\documentclass[aps,prl,twocolumn,groupedaddress,letterpaper,floatfix,showpacs]{revtex4}

\usepackage{graphicx}
\usepackage{subfigure}
\usepackage{rotating}
\usepackage{epstopdf}
\usepackage{amssymb}
\usepackage{epsfig}

\newcommand{\be}{ \begin{equation} }
\newcommand{\ee}{ \end{equation} }

\newcommand{\ignore}[1]{}
\newcommand{\bfomega}{\mbox{\boldmath $\omega$}}

\begin{document}



\title{Self-Replicating Three-Dimensional Vortices in Neutrally-Stable Stratified Rotating Shear Flows}
%

\author{Philip S. Marcus}
\author{Suyang Pei}
\author{Chung-Hsiang Jiang}
\author{Pedram Hassanzadeh}
\affiliation{Department of Mechanical Engineering, University of California, Berkeley, California, 94720, USA}


\date{\today}
\begin{abstract}
A previously unknown instability creates space-filling lattices of 3D vortices in linearly-stable, rotating, stratified shear flows. The instability starts from an easily-excited critical layer. The layer intensifies by drawing energy from the background shear and rolls-up into vortices that excite new critical layers and vortices. The vortices self-similarly replicate to create  lattices of turbulent vortices. The vortices persist for all time. This self-replication occurs in stratified Couette flows and in the dead zones of protoplanetary disks where it can de-stabilize Keplerian flows.
\end{abstract}

%
%
%
%
%
\maketitle


\paragraph {Introduction.}  For a protostar to  accrete gas from its protoplanetary disk (PPD) and form a star,  
the PPD must be
unstable and
transport angular momentum outward \cite{balbus98}.
This has led to efforts to find 
instabilities in 
PPDs and  other rotating flows that  satisfy Rayleigh's  criterion  for centrifugal stability, i.e., the
absolute value of angular momentum increases with increasing
radius  \cite{rayleigh}.
Numerical studies \cite{balbus96, shen2006}  of  PPDs and  experimental studies \cite{ji2006}
of  rotating flows  where the velocity obeys Rayleigh's  criterion confirm the stability of these flows
(although there are recent controversies \cite{paoletti2012,schartman2012,avila2012}).
In a PPD where the gas is sufficiently ionized to couple to magnetic fields,
the magneto-rotational instability (MRI) \cite{balbus98} operates. 
However, large regions of PPDs, known as {\it dead zones}, 
are too cool and  un-ionized to have  MRI.
Other instabilities  
\cite{Legal,lovelace1999} could de-stabilize a PPD, but they  require  unrealistic  boundaries
or continually-forced  perturbations.  Thus, star formation remains problematic. 

Here we report a new {\it finite-amplitude} instability in rotating, stratified, shearing 
flows in Cartesian or cylindrical geometries with velocities that would  satisfy Rayleigh's stability criterion if the
densities were constant (as assumed by Rayleigh). 
We examine rotating plane Couette flow, 
which is the canonical test for PPD stability. 
In previous studies using ideal gases  \cite{balbus96,balbus98,shen2006}, these  plane Couette flow PPD models were stable, but they were all 
initialized with no vertical density gradient
and no  vertical  gravity $g$. In contrast, here we include a stably-stratified initial density $\rho$ with $g \ne 0$ 
(as in a PPD). Previously, we observed, but did not understand, an instability in a PPD with an ideal gas and 
$g \ne 0$
\cite{barrancomarcus2005, marcus2013}. Thus,
to understand the instability, here we consider a Boussinesq fluid with constant $g$.
The 3D vortices found here are unique:
a vortex that grows from a single, small-volume, initial perturbation triggers a $1^{st}$-generation of 
vortices  nearby. This $1^{st}$-generation of vortices grows and  triggers a  $2^{nd}$-generation. The triggering of subsequent generations 
continues {\it ad infinitum}. The vortices do not advect in the cross-stream direction, but the front dividing the vortex-populated fluid 
from  the unperturbed fluid does. (Figs.~1 and~2.)  
Because the vortices grow large and  spawn new generations that  march  across the domain of  a {\it dead} zone, 
we refer to vortices that {\it self-replicate} to fill the domain as {\it zombie vortices}.  

%
\unitlength1mm{
\begin{figure}
\begin{picture}(83,51)
%
%
%
%
\put(1,1){\epsfig{figure=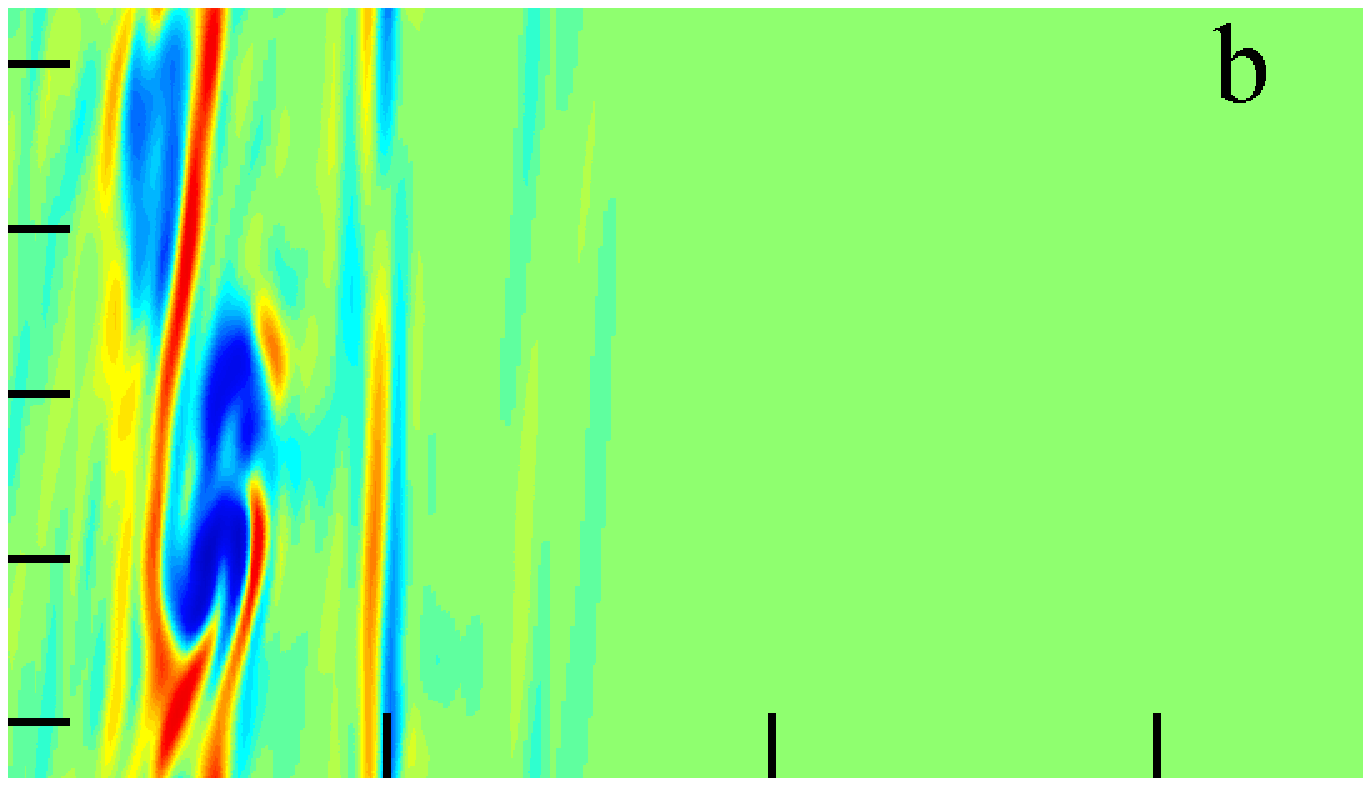,width=43mm}} 
\put(44,1){\epsfig{figure=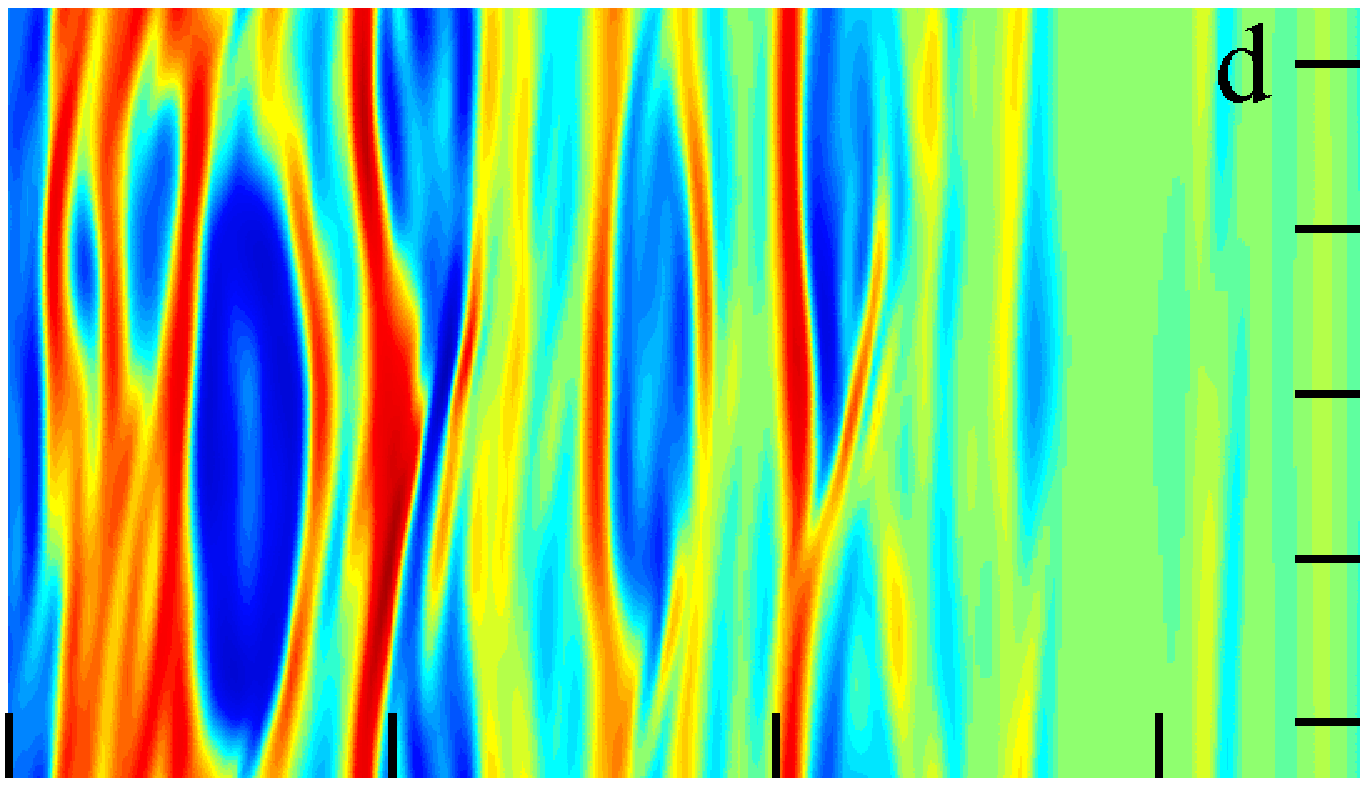,width=43mm}} 
\put(1,26){\epsfig{figure=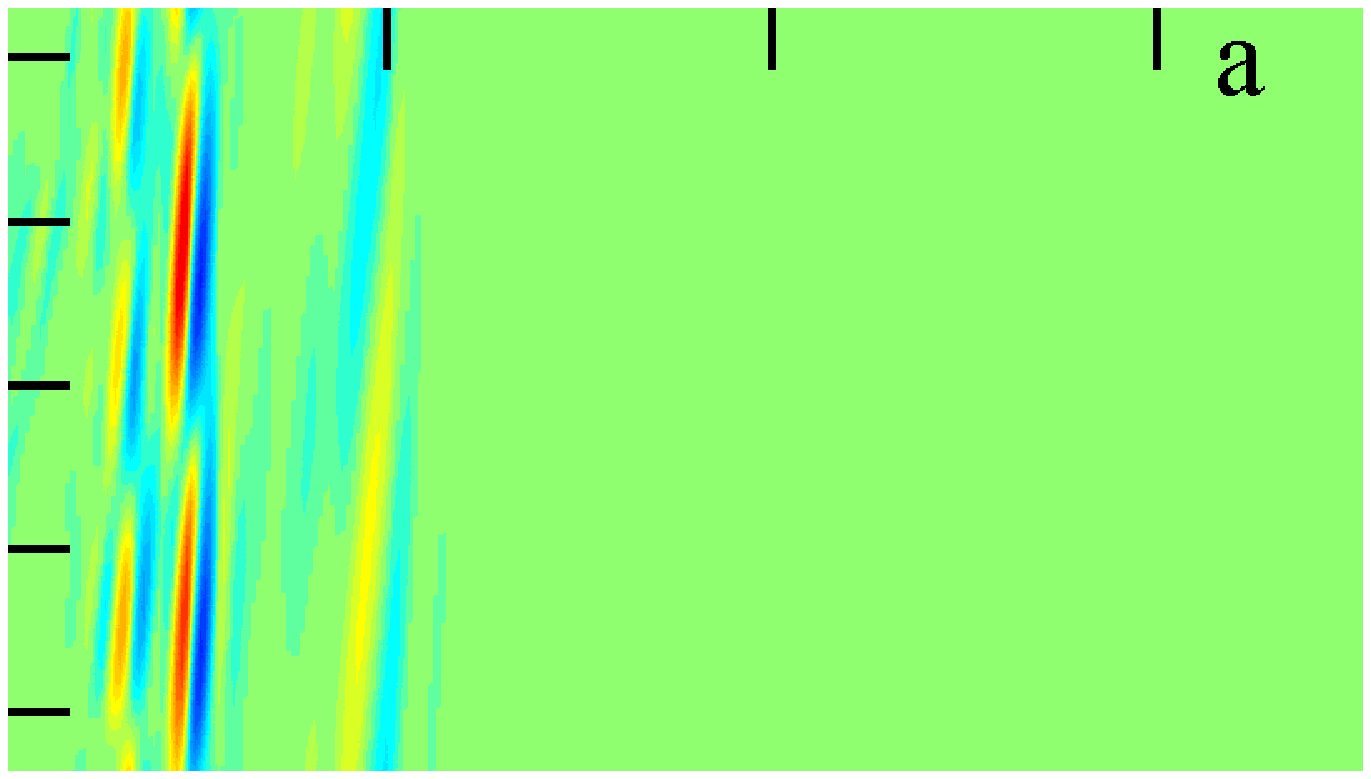,width=43mm}} 
\put(44,26){\epsfig{figure=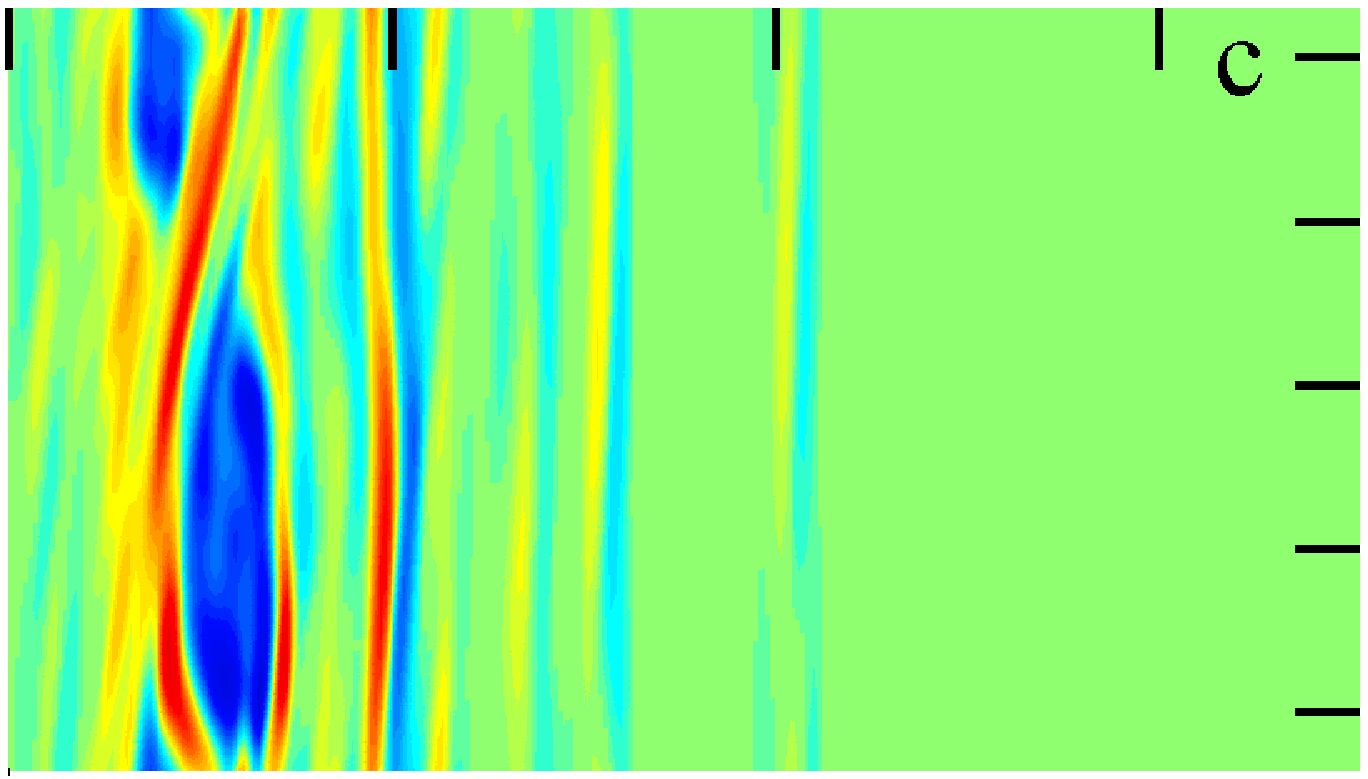,width=43mm}} 
\put(-0.4,47.5) {$2$}
\put(-0.4,42.25) {$1$}
\put(-0.4,37.25) {$0$}
\put(-2.9,32.25) {$-1$}
\put(-2.9,27) {$-2$}
\put(-0.4,22.75) {$2$}
\put(-0.4,17.5) {$1$}
\put(-0.4,12.5) {$0$}
\put(-2.9,7.25) {$-1$}
\put(-2.9,2) {$-2$}
\put(-3,25) {$y$}
\put(0.5,-1.2) {$0$}
\put(12.5,-1.2) {$1$}
\put(24.5,-1.2) {$2$}
\put(36.5,-1.2) {$3$}
\put(43.5,-1.2) {$0$}
\put(55.5,-1.2) {$1$}
\put(67.5,-1.2) {$2$}
\put(79.5,-1.2) {$3$}
\put(43,-3.75) {$x$}
\end{picture}
\renewcommand{\baselinestretch}{1.0}
\caption{\label{fig:1} 
    $\omega_z/f \equiv Ro$ of the  anticyclonic (blue) vortices and cyclonic (red) vortex layers in the $x$-$y$ plane. The initial 
perturbing vortex at the origin is above the plane shown here ($z=-0.4$). The first generation  zombie vortices form at $|x| \le 1$, 
and sweep outward in $x$. The Rossby number $Ro$ of these  vortices is $\sim$~$-0.2$. (The 
color is reddest at $\omega_z/f=0.2$, bluest at $\omega_z/f=-0.2$, and green at $\omega_z/f=0$).
$f/\bar{N} = 1$ and
    $\sigma/\bar{N} = - 3/4$.  The $x$-$y$ domain is 
    $|x| \le 4.7124$; $|y| \le 2.3562$, and is larger than shown. 
     Movies of Figs.~1 and~2 are online \cite{online}. 
        a) $t = 64/\bar{N}$. b) $t = 256/\bar{N}$. c) $t = 576/\bar{N}$. d) $t = 2240/\bar{N}$.
       See text for details.
}
\renewcommand{\baselinestretch}{2.0}
\end{figure}}
%

%
\unitlength1mm{
\begin{figure}
\begin{picture}(82,86)
%
%
%
%
\put(1,42){\epsfig{figure=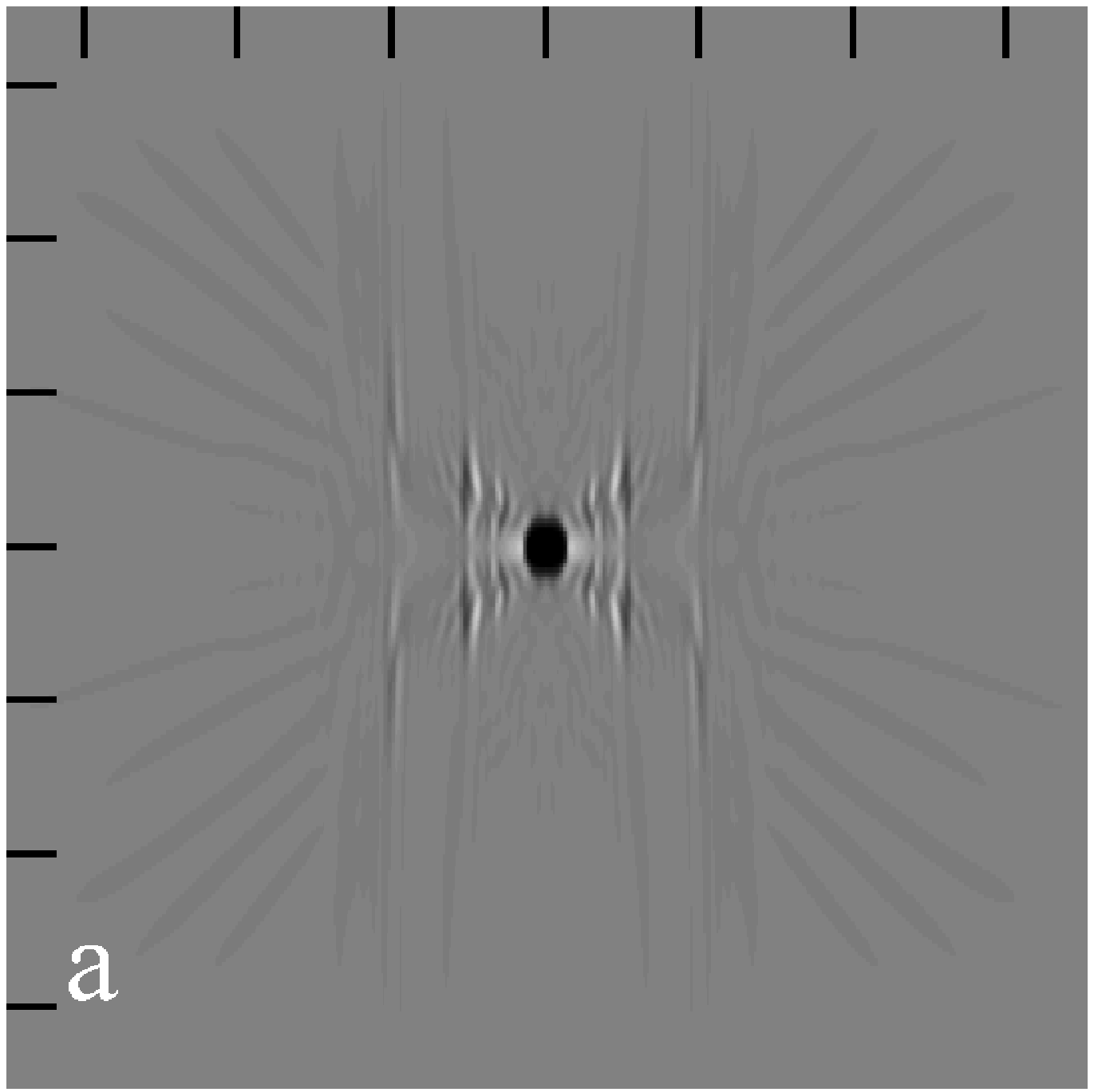,width=41.5mm}} 
\put(1,0){\epsfig{figure=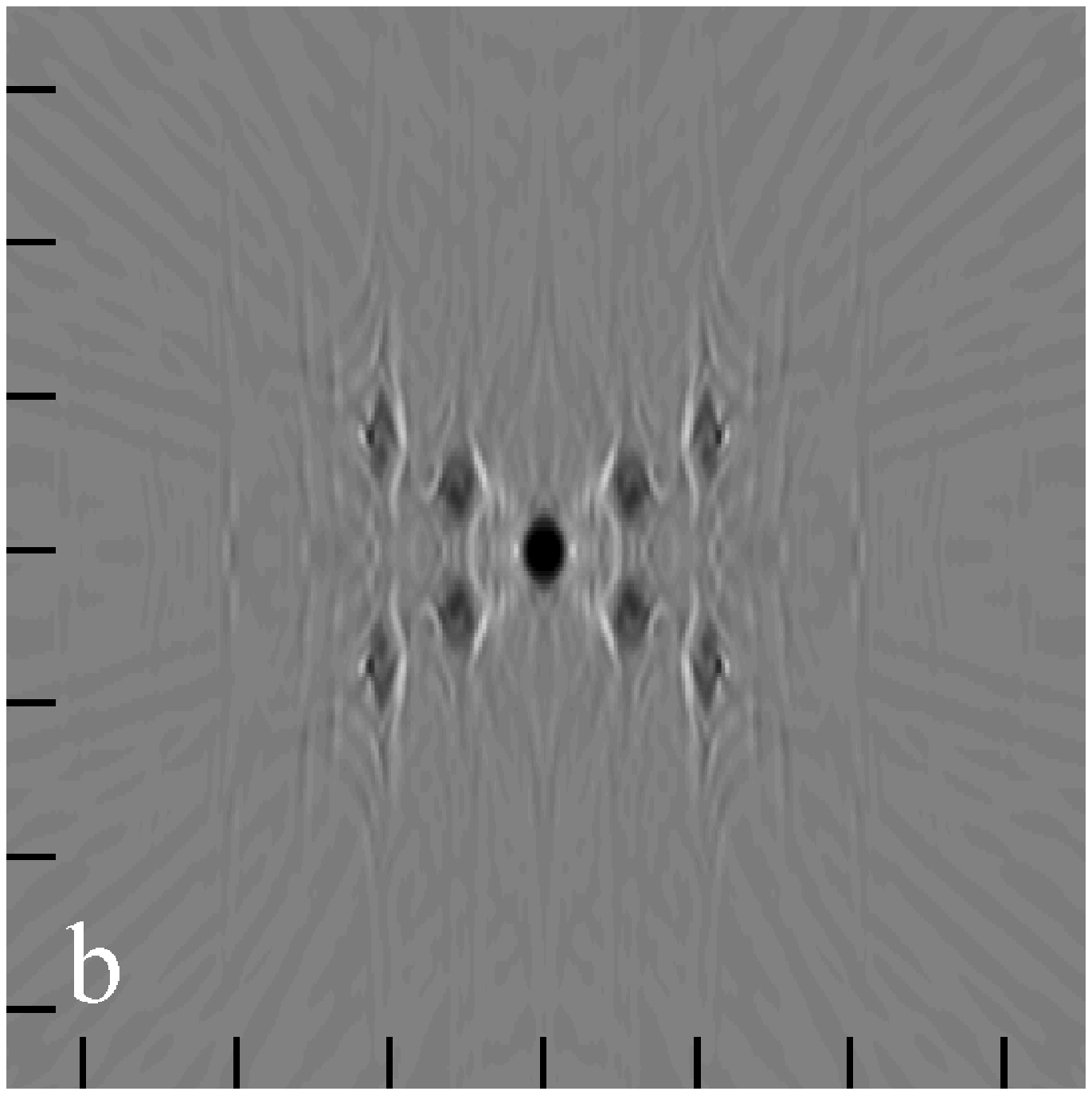,width=41.5mm}} 
\put(43,41.85){\epsfig{figure=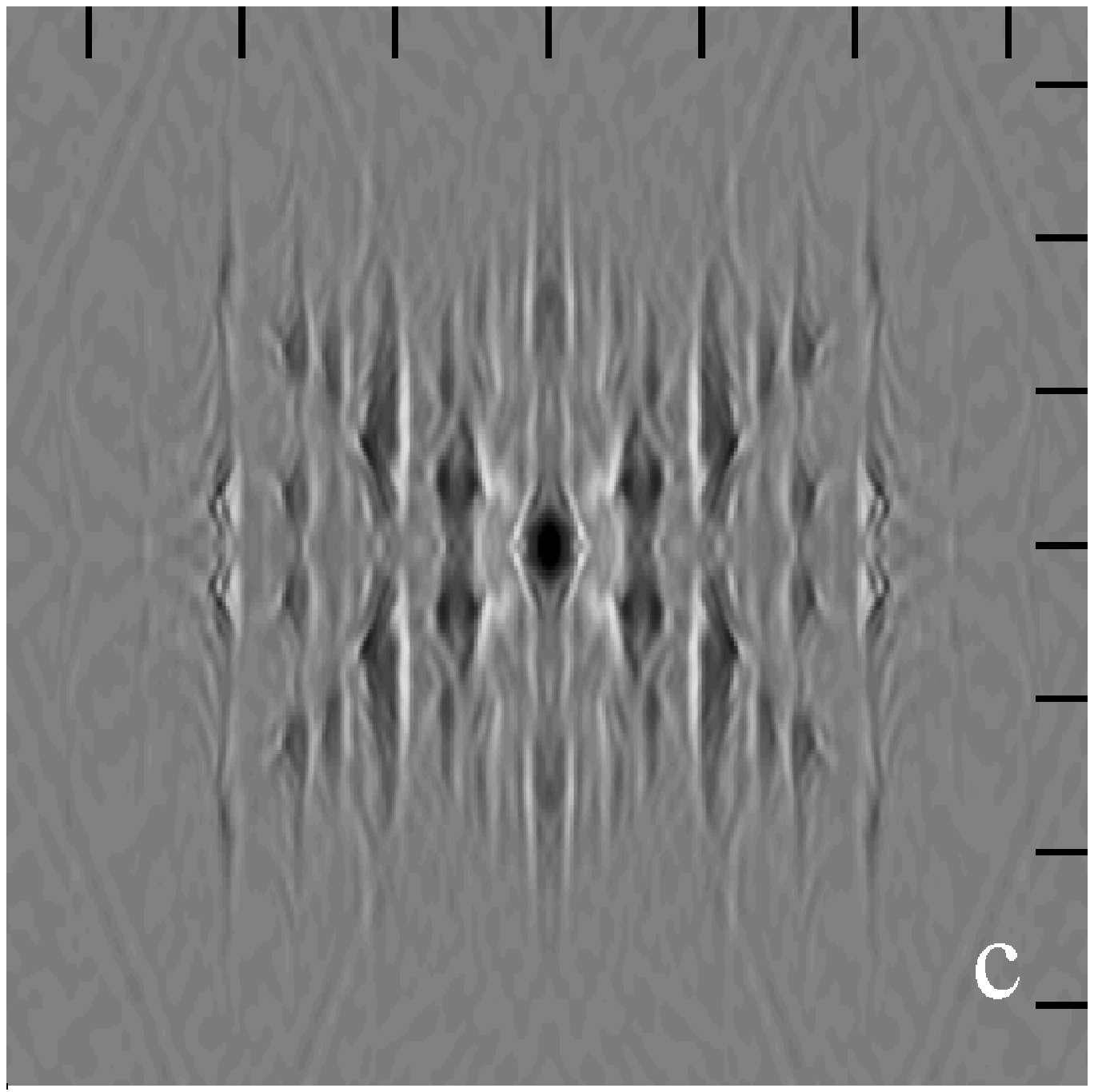,width=41.8mm}} 
\put(43,-0.15){\epsfig{figure=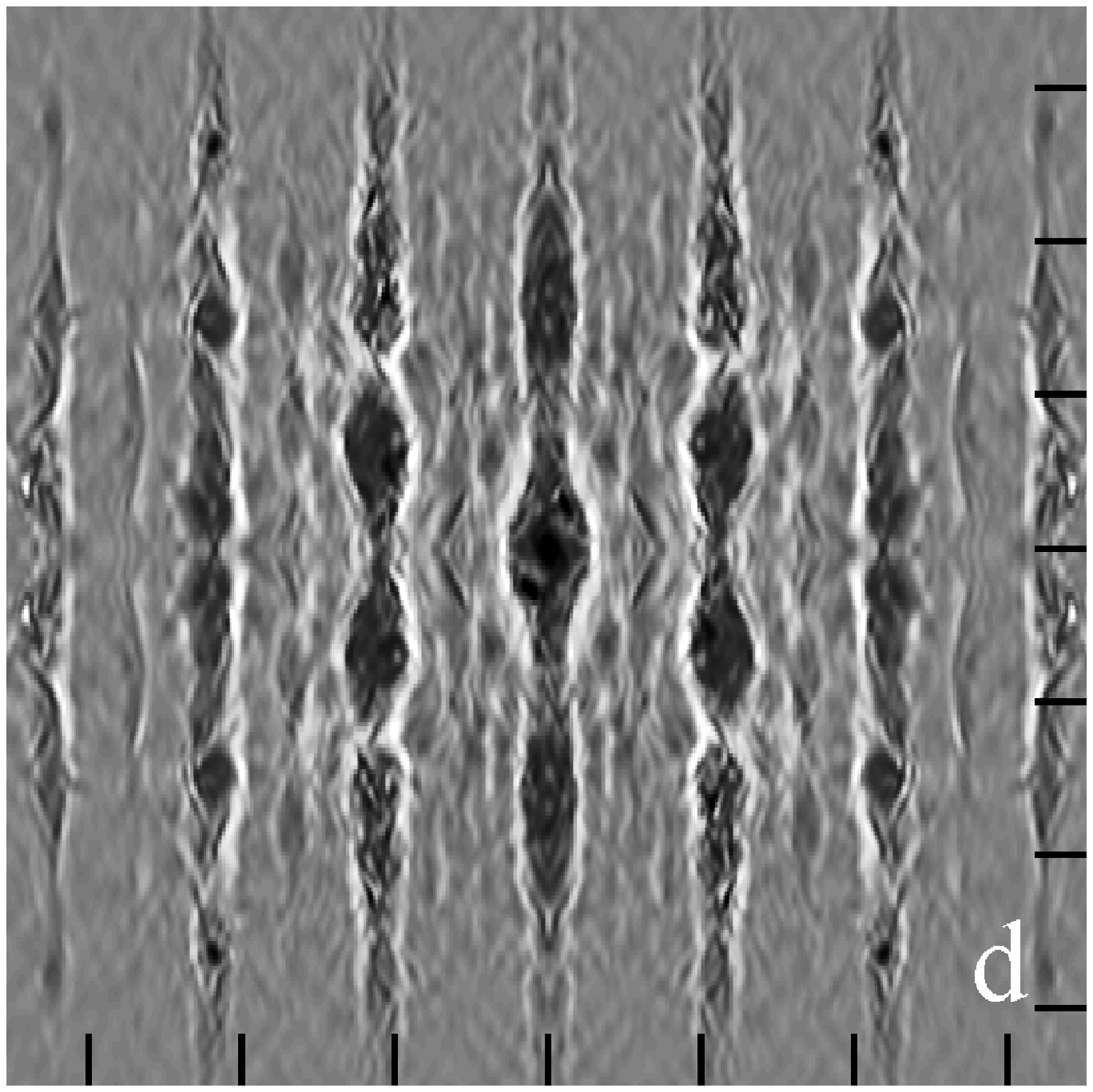,width=41.8mm}} 
\put(-0.5,79) {$3$}
\put(-0.5,73) {$2$}
\put(-0.5,67.25) {$1$}
\put(-0.5,61.5) {$0$}
\put(-3,55.75) {$-1$}
\put(-3,50) {$-2$}
\put(-3,44) {$-3$}
\put(-0.5,37.25) {$3$}
\put(-0.5,31.25) {$2$}
\put(-0.5,25.5) {$1$}
\put(-0.5,19.75) {$0$}
\put(-3,14) {$-1$}
\put(-3,8) {$-2$}
\put(-3,2) {$-3$}
\put(-3,41.25) {$z$}
\put(1,-2) {$-3$}
\put(7,-2) {$-2$}
\put(13,-2) {$-1$}
\put(21,-2) {$0$}
\put(26.75,-2) {$1$}
\put(32.75,-2) {$2$}
\put(38.5,-2) {$3$}
\put(43,-2) {$-3$}
\put(49,-2) {$-2$}
\put(55,-2) {$-1$}
\put(63,-2) {$0$}
\put(68.75,-2) {$1$}
\put(74.75,-2) {$2$}
\put(80.5,-2) {$3$}
\put(41.75,-3.75) {$x$}
\put(1,83.5) {$-3$}
\put(7,83.5) {$-2$}
\put(13,83.5) {$-1$}
\put(21,83.5) {$0$}
\put(26.75,83.5) {$1$}
\put(32.75,83.5) {$2$}
\put(38.5,83.5) {$3$}
\put(43,83.5) {$-3$}
\put(49,83.5) {$-2$}
\put(55,83.5) {$-1$}
\put(63,83.5) {$0$}
\put(68.75,83.5) {$1$}
\put(74.75,83.5) {$2$}
\put(80.5,83.5) {$3$}
\end{picture}
\renewcommand{\baselinestretch}{1.0}
\caption{\label{fig:2} 
Zombie vortices  sweep outward from the perturbing vortex at the origin in the $x$--$z$ plane (at $y=0$).  
Anticyclonic $\omega_z$ is black (darkest is $\omega_z/f=-0.2$)
and  cyclonic is white (lightest is $\omega_z/f=0.2$). This is the same flow as in Fig.~1. 
The domain has $|z| \le 4.7124$ and is larger than shown. 
a)  $t = 128/\bar{N}$. Critical layers  with $s=0$ and $|m|= 1$, 2, and~$3$ are visible. Diagonal lines 
are internal inertia-gravity waves with shear, not critical layers. 
b) $t = 480/\bar{N}$. $1^{st}$-generation vortices near $|x| = 1$ and $1/2$ have rolled-up from critical layers with $s=0$ and $|m|= 1$ and~$2$, respectively. 
c) $t = 1632/\bar{N}$. $2^{nd}$-generation $|m|=1$ vortices near $|x| =0$ and~$2$ were spawned from the $1^{st}$ generation vortices 
near $|x| = 1$. Another  $2^{nd}$-generation of $|m|=1$ vortices is
near $|x| \simeq 1/2$ and~$3/2$, which were  spawned by the $1^{st}$ generation near $|x| = 1/2$.
d)  $t = 3072/\bar{N}$. $1^{st}$, $2^{nd}$ and $3^{rd}$ generation vortices.
}
\renewcommand{\baselinestretch}{2.0}
\end{figure}}

The unperturbed velocity of plane Couette flow observed in a frame with angular velocity $\Omega \hat{\bf z} \equiv f/2 \hat{\bf z}$
is 
$\bar{{\bf v}} = \bar{V}(x)\,\, \hat{\bf y}$ with  $\bar{V}(x) \equiv \sigma x$, where $\sigma$ is the  uniform shear,
and $x$ and $y$ are the cross-stream and stream-wise coordinates. ``Hatted'' quantities are unit vectors.
The unperturbed density is 
$\bar{\rho}(z) = \rho_0 (1 - \bar{N}^2 z/g)$, where $\rho_0$ is constant  and $\bar{N} \equiv \sqrt{-g (d \bar{\rho}/dz)/\rho_0 }$ 
is the initial unperturbed Brunt-V\"ais\"al\"a frequency. In the rotating frame, the governing equations are
\begin{eqnarray}
\partial {\bf v}/\partial t &=& - ({\bf v} \cdot \nabla) {\bf v} - \frac{\nabla \Pi}{\rho_0} + f {\bf v} \times \hat{\bf z}
 -  \frac{(\rho - \rho_0) g}{\rho_0} \hat{{\bf z}} \label{v} \; \; \; \; \; \; \\
\partial \rho/\partial t &=& - ({\bf v} \cdot \nabla) \rho  \\
\nabla \cdot {\bf v} &=& 0, \label{div}
\end{eqnarray}
where $\Pi$ is the pressure head.
When Eqs.~(\ref{v}) --~(\ref{div}) are linearized about $\bar{V}(x)$ and $\bar{\rho}(z)$, the
eigenmodes are proportional to $e^{i(k_y y + k_z z - s t)}$.
When the initial density $\bar{\rho}$ is  stably-stratified or constant,
plane Couette flow  is neutrally linearly
stable (i.e., $s$ is real,
and eigenmodes neither grow nor decay).

\paragraph {Critical layers.} The eigen-equation for the eigenmodes of Eqs.~(\ref{v}) -- (\ref{div}) is a generalization of Rayleigh's equation 
\cite{dr81} and is a $2^{nd}$-order o.d.e. The coefficient of the highest-derivative term is
\begin{equation}
[\bar{V}(x) - s/k_y] \{[\bar{V}(x) - s/k_y]^2 - (\bar{N}/k_y)^2\}. \label{coefficient}
\end{equation}
Eigenmodes of an o.d.e. are singular
at locations $x^*$ where the  coefficient 
of the highest-derivative term 
is  zero. There they form
{\it critical layers} \cite {dr81}.
For fluids with constant density, critical layers obey $\bar{V}(x^*) = s/k_y$. 
We refer to these
as {\it barotropic critical layers}.
For $\bar{N} \ne 0$, expression~(\ref{coefficient}) shows that there are eigenmodes with 
barotropic critical layers, but our computations show that they are difficult to excite and never form vortices.
However,
there is another class of eigenmodes with critical layers; they have  $\bar{V}(x^*) - s/k_y \pm \bar{N}/k_y = 0$, and we call them
{\it baroclinic critical layers}.
Weak baroclinic critical layers were  shown to exist in non-rotating, stratified flows
\cite{boulanger2007}, but we believe that this is the first study of these layers in flows with  $f$, $\bar{N}$ and $|\sigma|$ of the same order 
(as near the mid-plane of a PPD).
With anticyclonic shear ($ f\sigma < 0$), as in a PPD, all of our calculations with $\bar{N} \simeq f \simeq |\sigma|$ fill the domain with zombie vortices when 
the initial finite-amplitude perturbation is sufficiently large (see below). To verify our computations,  flows were computed with two independent 
codes.
At  the  $x$
boundaries, one code enforced an outward-going wave condition, and the other used the shearing sheet approximation \cite{barrancomarcus2006}. 
The codes produced similar results.

We show that the  new finite-amplitude instability works by first creating
large-amplitude vortex layers at the critical layers. The curl of
Eq.~(\ref{v}) gives 
\begin{equation}
\partial {\omega_z}/\partial t = - ({\bf v} \cdot \nabla) \omega_z + (\bfomega \cdot \nabla) v_z  + (f + \sigma)  (\partial v_z/\partial z), \label{omega}
\end{equation}
where $\bfomega$ is the {\it relative} vorticity 
$\bfomega \equiv \nabla \times ({\bf v} - \bar{V}(x) \, \hat{\bf y})$. Vortex layers form  
at  baroclinic critical layers because the $z$-component
of the velocity $v_z$ of the neutrally stable eigenmode is singular there.
Equation~(\ref{omega}) shows that the generalized Coriolis term
$(f + \sigma) (\partial v_z/\partial z)$ creates $\omega_z$.
Within the baroclinic critical layer, the singular $\partial v_z/\partial z$ is nearly anti-symmetric about
$x=x^*$; on one side of the  layer $v_z \rightarrow \infty$, and on the other $v_z \rightarrow - \infty$; 
thus, the last term in Eq.~(\ref{omega}) creates
a large-amplitude vortex layer centered at $x^*$ made of dipolar segments with one side cyclonic  ($\omega_z f > 0$) and the other 
anticyclonic  ($\omega_z f < 0$) (c.f., Fig.~1(a)). Barotropic critical layers do not form vortex layers; although their eigenmodes' 
$v_y$ is singular, $v_z$ is everywhere
finite. 
From this point on, we use non-dimensional units
with the units of time $1/\bar{N}$ and length  $|(L\bar{N})/(2 \pi \sigma)|$, where $L$ is the periodicity length in $y$.
Thus, $k_y$ in expression~(\ref{coefficient}) is $2 \pi m/L$, 
where $m$ is an integer. 
Baroclinic
critical layers have  $k_y \ne 0$, and expression~(\ref{coefficient}) shows that they are at:
\begin{equation}
x^* = - (s \pm 1)/m. \label{clinic}
\end{equation}
Equations~(1)~--~(3) and their boundary conditions are invariant under translations in $y$ and $z$, and also under translation in 
$x$ by  $\delta$ when accompanied by a stream-wise boost in velocity of $\sigma \delta$. The latter symmetry is
{\it shift-and-boost symmetry}, c.f., \cite{goldreichlyndenbell, marcuspress} and is the basis of  the {\it shearing sheet} boundary 
conditions  \cite{barrancomarcus2006, balbus98}.
Due to the shift-and-boost symmetry, the origin of the $x$-axis is not unique, so
Eq.~(\ref{clinic}) has the following meaning: $x^*$ is the cross-stream distance  between a perturbation and 
the location of the baroclinic critical layer that it excites.

Many types of perturbations
create zombie vortices. Most  relevant to PPDs is a Kolmogorov spectrum of noise where the velocity and Rossby number $Ro \equiv \omega_z/f$ 
of the initial eddies scale respectively as 
${\it l}^{1/3}$ and  ${\it l}^{-2/3}$, where ${\it l}$ is the eddy diameter. 
The smallest eddies have the largest vorticity and $Ro$. In calculations with  $\sigma/f=-3/4$  and $0.5 \le \bar{N}/f \le 1$ (the regions
we explored in a PPD \cite{barrancomarcus2005}), 
regardless of how small
we make the amplitude of the initial Kolmogorov energy spectrum, if the spatial resolution is sufficient, the smallest eddies have a  sufficiently large 
$|Ro|$ to trigger the instability and create zombie vortices. The vortices eventually  fill the domain, such that 
at late times the volume they occupy  is of order of the domain's volume. To better understand zombie vortex formation and replication, we simulated flows
with   $\sigma/f=-3/4$ and $0.5 \le \bar{N}/f \le 1$ initialized with a single ``shielded'' \cite{hassanzadeh2012} anticyclone at the origin. These initial conditions produced 
flows filled with zombie vortices 
with $-0.35 < Ro < -0.15$ when the initial anticyclone had $|Ro| \gtrsim 0.2$.
Figs.~1 and~2 illustrate the case where the initial anticyclone has $Ro=-0.31$ (as in the PPD where we first observed zombie vortices \cite{barrancomarcus2005})
and volume  $\sim$~$10^{-4}$ of the  domain.  
The velocity perturbation due to the  initial vortex is significant only  near the origin and is
small, $\sim$~$10^{-2} \sigma L_x$, where $L_x$ is the domain size in $x$.
(Velocity perturbations in PPD studies are considered small when they are less than $\sim$~$0.1 \sigma L_x$ \cite{balbus96}.)  
Our initial vortex is in quasi-equilibrium as in \cite{barrancomarcus2005} such that Eqs.~(1) and~(3), but not~(2), are in approximate steady equilibrium. 
The initial density perturbation is confined to the initial vortex. Eq.~(2) allows $\rho$ and $N(x,y,z,t)$ to change.
Figure~1 shows  $\omega_z$ in an $x$--$y$ plane. 
The perturbing vortex is nearly steady, so it  excites critical layers with frequencies $s=0$. 
Thus, Eq.~(\ref{clinic}) shows that the critical layers are at $|x^*| = 1/|m|$ with
no critical layers at $|x| > 1$. 
Figure~1(a) shows vortex layers at these critical layers:
$\omega_z$ appears at $x =1/|m|$ as $|m|$ 
segments of dipolar stripes aligned in the stream-wise $y$ 
direction for $|m| =1$, 2 and 3.
A  Fourier analysis shows that the stripes
have $s=0$. 
We previously showed \cite{marcusjfm, marcusn} that in shear flows with $f \sigma < 0$,
cyclonic vortex layers aligned in the stream-wise direction are stable, whereas 
anticyclonic layers are unstable, roll-up into discrete anticyclones, 
and  merge to form one large anticyclone. 
This behavior is seen in Fig.~1(b).
The 
anticyclonic vorticity at $x=1/3$  has  rolled up and merged into  a single anticyclone (near $y=1.5$).
The 
anticyclonic vorticity at $x=1/2$  has  rolled up into an anticyclone near $y=-0.5$.
In contrast, the cyclonic $\omega_z$ near $x=1/2$ has formed a continuous, meandering filament. At later times (Fig.~1(c))
the anticyclones near  $x=1/3$  (and near $y=2$) and  near  $x=1/2$  (and near $y=-1$) have become larger. 
Figures~1(c) and~1(d) show critical layers and vortices at $|x| > 1$, which cannot be created by  perturbations at the origin. 
The layers at $|x| > 1$ are due to the  self-replication of  $1^{st}$-generation vortices at $|x| \le 1$.
A vortex at {\it any} location will  excite critical layers in a manner exactly like the original perturbing
vortex due to the
{shift-and-boost} symmetry (and will have $s=0$ when viewed in the frame moving with the perturbing vortex). 
Figure~1(c) shows  $2^{nd}$-generation critical layers at $x=4/3$, $3/2$, $2$, and $2/3$ all with $|m|=1$ and excited by  $1^{st}$-generation vortices at
$x = 1/3$, $1/2$, $1$, and $-1/3$, respectively.
Figure~1(d) shows  $3^{rd}$-generation critical layers at $2 < x \le 3$, and  $4^{th}$-generation critical layers forming at $x>3$.
At later times the vortices from $|m|=1$ critical layers dominate (Fig.~2(d)).
At very late times, the vortices have  cross-stream diameters   of order unity.
(See below.) Within each zombie vortex the density  mixes so that it is in accord with 
its near hydrostatic and geo-cyclostrophic equilibrium (c.f., \cite{hassanzadeh2012}). 
However, there is horizontal, but very little vertical,  mixing of density outside the vortices, 
so the background vertical density stratification and $N$ remain within $1\%$ of their initial unperturbed values.
The lack of vertical mixing, despite strong horizontal mixing, was seen in our earlier simulations  \cite{barrancomarcus2005}  and  laboratory experiments \cite{aubert2012} of vortices in rotating, stratified flows.

Figure~2 shows the flow in Fig.~1 viewed in the $x$--$z$ plane and illustrates our main result:  
at late times the domain   fills with anticyclones. 
Because the initial flow is homogeneous with uniform $\sigma$ and $\bar{N}$, the vortices form a regular lattice despite the flow's turbulence.
As time progresses in Fig.~2, the vortex population spreads out from the perturbing vortex at the origin. At early times (Fig.~2(a)) the flow has 
$1^{st}$-generation  critical layers,
with $|m| =1$, 2, and 3 being most apparent. In this first generation, and all subsequent generations, a vortex perturbs the flow and creates four new prominent 
vortices at its  $|m|=1$ critical layers
at locations in $x$ that are $\pm l_x$ distant
from itself and at locations in $z$ that are $\pm l_z$ distant from itself. 
($l_x$ is physically set by, and equal to, the distance in $x$ from a perturbing vortex to the anticyclonic piece of the vortex layer formed by its $|m| = 1$ critical layer; this distance is slightly  greater than unity.)
The 
$2^{nd}$-generation $m=1$ critical layers  created by the $1^{st}$-generation  vortices with $|m|=1$, 2, and 3 
are faintly visible in Fig.~2(b) and much more so in  Fig.~2(c). At later times (Fig.~2(d)), the $|m|=1$ 
vortices 
descended from the 
$1^{st}$-generation $|m|=1$ vortices dominate and form
a lattice of zombie vortices located at
[$x = 2 n \,\, l_x, z= 2 j \,\, l_z$] and at [$x = (2 n + 1) l_x, z= (2 j +1)  l_z$], for all integers $n$ and $j$.

The characteristic $|Ro|$ of late-time zombie vortices in Figs.~1 and~2 is $\sim$~0.2, consistent with zombie vortices in flows initialized with noise.
After a  vortex forms, its $|Ro|$ intensifies to its approximate peak value  within a few 
of its turn-around times, and it remains near that  value indefinitely. Based on several numerical experiments, it appears that the late-time values
of $|Ro|$ depend on the  parameters, $\bar{N}$, $f$ and $\sigma$ rather than on  properties of the initial perturbation.
To examine the energy of the vortices and discover its source, we decomposed the 
flow's energy into two orthogonal parts: (1) the {\it zonal} component consisting of the kinetic energy of the 
stream-wise velocity component with Fourier modes $k_z = k_y = 0$ (i.e., the background shearing flow); and (2) the  {\it non-zonal} component  consisting of 
everything else, including the potential energy $g \int z (\rho - \bar{\rho}) \,\, (d \, {\rm volume})$.
If the initial flow were  unperturbed,
then the initial energy would be all zonal.
In the flow in Figs.~1 and~2, there is a small  initial non-zonal
component due to the initial vortex at the origin.
At later times, the non-zonal component represents the energy of the initial vortex and the zombie 
vortices (and turbulence and waves). The non-zonal energy initially increases super-exponentially 
for $0 \le t \lesssim 190$, increasing to $\sim$~$15$ times its
initial value. Then, the non-zonal energy  increases approximately exponentially 
with an e-folding time of $\sim$~860, such that at $t = 3072$ in Fig.~2(d), the non-zonal energy is more than 400 times its initial value.
The energy increase in the non-zonal component is supplied by the zonal energy. 
The exponential growth of the non-zonal energy is 
due to the fact that vortices in the vortex-populated region grow exponentially in size, and not due to a long-term exponential increase of the velocity of each zombie vortex. 
Therefore, the non-zonal energy must plateau once the vortices fill the domain. 
If the self-replication were self-similar, we would expect the perimeter of the front between the vortex-populated flow and unperturbed flow in each $x$-$z$ plane to grow as $t$ and the number of vortices to increase as $t^2$, which is consistent with our calculations.

\paragraph {Discussion.} We have shown that linearly, neutrally stable plane Couette flow becomes finite-amplitude unstable when it is vertically stably-stratified.
In the example here,  baroclinic
critical layers are excited by a small vortex, but our calculations show that a variety of small-volume, small-energy perturbations cause
critical layers to grow  and roll-up into large-volume, large-energy vortices. 
In general, this instability self-replicates with each new vortex exciting new  layers that roll-up until the
domain fills with compact 3D (i.e., not Taylor columns)  vortices. The robustness of zombie vortices 
is evident from the fact that  they survive indefinitely even though they are embedded in a turbulent flow at late times.
They survive by drawing energy from the background  shear flow.
For constant $\bar{N}$ and $\sigma$, the unperturbed flow is homogeneous, and  vortex
self-replication is self-similar with zombie vortices forming a regular lattice. 
The regularity of the lattice allows for reinforcement: each vortex  re-excites four other vortices in the lattice, and each vortex in the
lattice is continually re-excited by four other vortices.
Zombie vortices occur frequently in our  simulations of 
Boussinesq and
compressible fluids,
so they pose a paradox: if they are so common, why have  they not been reported earlier? We believe there are three reasons: 
(1) instabilities have not been systematically sought in stratified
Couette flows \cite{Legal}; (2) with few exceptions \cite{zahn}, stability 
studies of ideal gases in PPDs were carried out  with no initial vertical stratification \cite{balbus96, shen2006}; and (3) 
the necessary spatial resolution to compute critical layers  is lacking in many calculations.
Zombie vortices occur in our  calculations of the dead zones of protoplanetary disks  \cite{barrancomarcus2005}, which suggest that they may have 
an important role in star
and planet formation. In addition, zombie vortices should be observable in laboratory circular Couette flows with stratified salt water 
for parameter values  where the flow is linearly stable with respect to centrifugal instability \cite{dr81}, SRI \cite{molemaker2001,Legal} and other instabilities 
\cite{ledizesbillant2009}.

\begin{acknowledgments}
We thank NSF-XSEDE, NASA-HEC, NASA-PATM, NSF-ATI and NSF-AST for support. 
\end{acknowledgments}

\end{document}